\documentstyle[prl,aps,epsfig,multicol]{revtex}
\begin{document}
%\preprint{IMSc-2003/06/16}
\def\be{\begin{equation}}
\def\ee{\end{equation}}
\def\bearr{\begin{eqnarray}}
\def\eearr{\end{eqnarray}}
\def\tc{$T_c~$}
\def\tcl{$T_c^{1*}~$}
\def\c2{ CuO$_2~$}
\def\ruo{ RuO$_2~$}
\def\lsco{LSCO~}
\def\bi{bI-2201~}
\def\tl{Tl-2201~}
\def\hg{Hg-1201~}
\def\sro{$Sr_2 Ru O_4$~}
\def\rc{$RuSr_2Gd Cu_2 O_8$~}
\def\mgb{$MgB_2$~}
\def\pz{$p_z$~}
\def\ppi{$p\pi$~}
\def\sqo{$S(q,\omega)$~}
\def\tperp{$t_{\perp}$~}
\def\cob{$CoO_2$~}
\def\nxcob{$Na_x CoO_2.yH_2O$~}
\def\ncob{$Na_{0.5} CoO_2$~}
\def\half{$\frac{1}{2}$~}
\def\nycob{$A_xCoO_{2+\delta}$~}
\def\naxcob{$Na_xCoO_2$~}
\def\wat{$H_2O$~}
\def\na{$Na$~}
\def\ewat{$\epsilon_{\rm H_2O}$~}

\title{How Ice enables Superconductivity in \nxcob by melting charge order:\\
Possibility of novel Electric Field Effects}

\author{ G. Baskaran\\
The Institute of Mathematical Sciences\\
C.I.T. Campus, Chennai 600 113, India }

%\date{today}

\maketitle

\begin{abstract}

Charge ordering in doped \cob planes near the commensurate fillings
$x = \frac{1}{4}$ and $\frac{1}{3}$ are considered for \nxcob and 
suggested to be competitors to superconductivity, leading to 
the experimentally seen narrow superconducting dome bounded by 
commensurate doping: $\frac{1}{4}<x< \frac{1}{3}$. 
Intercalated hydrogen bonded \wat 
network, by its enhanced dielectric constant, screen and frustrate 
local {\em charge order condensation energy} and replace a generic 
`charge glass order' by superconductivity in the dome.  An access 
to superconductivity and charge order, available through the new 
water channel, is used to predict novel effects such as 
`Electrical Modulation of Superconductivity' and `Electroresistance
Effect'.
\end{abstract}

%\pacs{PACS number: 13.20.He, 11.30.Er, 12.15Hh, 11.80Et}

%\maketitle

\begin{multicols}{2}[]

Discovery of superconductivity in \nxcob by Takada and 
collaborators\cite{takada} have opened the possibility of realizing
unconventional superconductivity and novel quantum states in 2D arising 
from strong electron correlations in doped \cob layers. Water of a 
right proportion ($y \approx \frac{4}{3}$) seems absolutely 
necessary\cite{cava1,cava2,chu,jin} for stabilizing 
superconductivity, suggesting $H_2O$'s critical role. While water does 
wonders in nature, its key role here is some what puzzling. Elucidating 
its role in this unusual superconductor is an important 
task from material science and physics point of view.  This is what 
the present paper attempts using phenomenological and theoretical
considerations.

Enthused by the remarkable discovery of superconductivity in \nxcob,
the present author\cite{gbcob} and others\cite{rvbcob}
have suggested a single band t-J model
as an appropriate model to understand superconductivity and low energy 
electronic phenomena. A phase diagram has been suggested using ideas 
of resonating valence bond (RVB) theory developed for 
cuprates\cite{pwagbrvb}. A recent 
experiment\cite{cava2} which shows superconductivity in a 
rather narrow range 
of doping $\frac{1}{4}<x< \frac{1}{3}$, than predicted by RVB theories, 
suggest that there are perhaps left out interactions and consequent 
competing phases which make a simple t-J model valid only for 
limited range of $x$. The situation is not unusual - even in cuprates 
a simple t-J modeling is strictly valid only in the neighborhood of 
optimal doping. Charge order phenomenon\cite{stripe,stripe1} is known 
in cuprate superconductors; and it has been suggested to  
compete\cite{gbstripe} with superconductivity.

Doped \cob, compared to $CuO_2$ layers of high \tc cuprates, has a 
narrower conduction band and a less polarizable valence band of oxygen. 
Consequently, short range coulomb repulsions among carriers  
are screened less. This is likely to stabilize a variety of frustrated 
charge ordering in the triangular lattice, as we discuss in this 
paper. NMR result of Ray et al.\cite{susc} indeed provides a first evidence 
for charge freezing in \naxcob family ($x = \frac{1}{2}$), \ncob below 
about $T \approx 300 K$, a large temperature scale. 

We estimate and include unscreened short range coulomb interactions 
in the t-J model for the study of unhydrated \naxcob. 
We show that in addition to superconductivity, charge ordering in the 
narrow conduction band of \cob layer is a major instability, for a 
range of higher doping than suspected (Wen et al. in ref 7). 
As the unscreened
near neighbor coulomb interactions are large and comparable to the
band width, the characteristic charge order temperatures are
$T_{\rm ch} \sim 400 K$. 

Fortunately, \wat, in hydrated \nxcob, makes t-J modeling valid 
for a range of doping. For reasons which we elaborate in the 
present paper, hydrogen bonded \wat dipoles of the ice layers screen and 
frustrate {\em charge order condensation energy}. That is,
they {\em effectively 
screen out short range repulsions}, and enable the physics of a simple 
t-J model to be realized in a narrow range of doping, 
$\frac{1}{4}<x< \frac{1}{3}$, as superconductivity. 

We discuss few important charge ordered states at commensurate 
fillings, $x = \frac{1}{4}$ and $ \frac{1}{3}$, which we believe are
competitors to the experimentally observed superconductivity, in the 
range $\frac{1}{4} < x < \frac{1}{3}$. These reference charge ordered 
states are strongly frustrated by the random potential from the 
neighboring $Na$ layers, resulting in a glassy phase in the region
$\frac{1}{4} < x < \frac{1}{3}$ and beyond. The charge glass phase
is likely to be an anomalous metal, very much like the spin gap phase
in cuprates, where there are local charge order activities
at low frequency scales. 

We estimate the enhancement of the background static dielectric 
constant at short distance due to hydrogen bonding in the \wat layers.
We find that this screening is sufficient to reduce the large charge
order transition temperature down to $\sim 1 K$ and allow 
superconductivity to emerge. Strong commensurability effects and
the associated short range charge order reduce superconducting \tc 
considerably as we approach the commensurate ends $ x = \frac{1}{4} $
and $\frac{1}{3}$.

As \wat stabilize superconductivity and discourages charge
order, we have a new access to the electronic phases of \cob layer
through water. This leads to the possibility of some novel effects: i) 
`Electrical Modulation of Superconductivity' 
by external electric field or microwave radiation and 
ii) `Electroresistance Effect' in the normal state.
We estimate that voltages $\sim 500 V$, applied capacitively 
to thin films of \nxcob of thickness $\sim~1$ micron will 
orient the water dipoles and reduce the short distance 
dielectric screening, resulting in stabilization of charge glass order
phase and destabilization of superconductivity. This interesting 
switching effect may have device potential.

Recently we modeled the low energy physics of doped \cob using 
a t-J model and discussed an RVB scenario for superconductivity including
a PT violating $d_1+id_2$ wave superconductivity and a $p_1+ip_2$ wave 
superconductivity at a higher doping. To study charge order we must 
include some leading short distance carrier-carrier and carrier-$Na$-ion 
screened coulomb interaction: 
\bearr
H_{{\rm tJV}} =
 -t  \sum_{\langle ij\rangle} C^{\dagger}_{i\sigma} C^{}_{j\sigma}
+ H.c. +  J \sum_{\langle ij\rangle} 
({\bf S}_{i }\cdot{\bf S}_{j} - \frac{1}{4}n_i n_j) \nonumber \\
\sum_{ij} V_{ij} (n_i-1)( n_j-1) + \sum_{i}\epsilon_i (n_i -1)
\eearr
Here $C$'s and ${\bf S}$'s are the electron and spin operators.
As we have an electron doped system we have the `zero occupancy' 
constraint $\sum_{\sigma} n^{}_{i\sigma} \neq 0 $ at every site i. 

Recall that doubly occupied $Co^{3+}$ sites carry a charge 
$-e$ with reference to the neutral \cob layer and $V_{ij}$ is the 
screened coulomb repulsion between them. We have ignored the small two body 
off-diagonal coulomb interaction terms. For practical purposes only 
the nearest and next nearest neighbor terms $ V_1 \approx  
\frac{e^2}{\varepsilon_{ab} R_{nn}} e^{-\frac{R_{nn}}{\lambda_{ab}}}$
and $ V_2 \approx  \frac{e^2}{\varepsilon_{ab} R_{nnn}}
e^{-\frac{R_{nnn}}{\lambda_{ab}}}$ are important.
Here $\varepsilon_{ab} \approx \varepsilon_O + \varepsilon_{\rm {H_2O}} $ 
represents the short distance dielectric screening arising from the filled 
oxygen bands of \cob layers and $H_2O$ layers in \nxcob. And 
$\lambda_{ab} \approx Co-Co$ distance is the Thomas Fermi screening length 
for our tight binding metallic layer. The random site energy $\epsilon_i$ 
of charge degree of freedom represents the screened 
coulomb attraction from neighboring $Na^+$ ions.

Electronic structure calculations\cite{singh} give a value of 
$t \approx -0.1~eV$ for the conduction band of the \cob layer. 
We estimate $V_1$ and $V_2$ for \naxcob, 
the non-hydrated case. The dielectric constants of oxides of $Fe$ and $Ni$ 
that flank $Co$ in the periodic table are $\sim 4$ to $12$. We assume a 
background (short distance) static dielectric constant of 
$\varepsilon_{ab} \approx 8$ for our \cob layer. Recall that in 
cuprates the background $\epsilon$ is large $\sim 20$, in view of 
a more polarizable octahedral oxygen network; in \cob the oxygen filled
band is less polarizable and relatively deep below the fermi level.
Using this 
dielectric constant and values of Co-Co distances in \ncob 
we get $V_1 \approx 0.8~eV$ and $V_2 \approx 0.4~eV$. The mean square
fluctuation of the carrier site energy due to disordered $Na$-ions
is ${\sqrt{\langle \delta \epsilon_i^2 \rangle}} \approx 0.2~eV $ 

In the absence of hopping, the dopant carriers $Co^{3+}$ will order 
classically and undergo order-disorder transition at a fairly high 
temperature $k_B T_{\rm ch}(\rm classical)
\approx 2{\bar V}\sim 10^3~K$; here 
${\bar V} \equiv \frac{1}{2}(V_1 + V_2)$ is a mean short distance 
repulsion. However the electron dynamics reduce 
$T_{\rm ch}(\rm classical)$ considerably. To estimate this reduction we
perform a mean field analysis of the t-J-V model for a CDW order, 
pretending that an unfrustrated charge order arises from nesting 
instability for \ncob. This gives us a BCS like expression for \tc:
\be
k_BT_{\em ch}  
\approx \epsilon_F e^{-\frac{1}{{\bar V}\rho_o}} 
\ee
Here $\rho_o$ is
a fermi sea averaged particle-hole density of states corresponding to 
the ordering wave vector. Substituting $\epsilon_F \approx 0.5~eV$, 
$\rho_0 \approx \frac{1}{2\epsilon_F}$ and ${\bar V} \approx 0.4$, we get  
 $T_{\rm ch} \approx 480 K$. Frustration on the triangular lattice at 
 half filling
and disorder effect from $Na$ ions will further reduce this. Thus we get
a charge order temperature in the right range, 
$T_{\rm ch}(NMR) \approx 300~K$, seen in NMR, 

\begin{figure}[h]
\epsfxsize 5.5cm
\centerline {\epsfbox{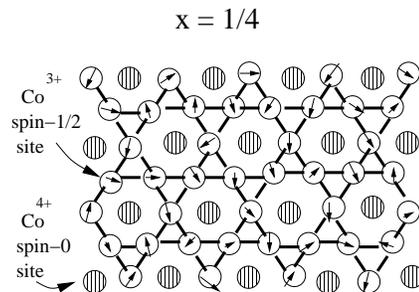}}
\caption{`Classical' charge order for $x = \frac{1}{4}$.
Spin-0, charge $-e$ carriers ($Co^{3+}$) form a triangular lattice. 
Neutral sites with spin-\half moments ($Co^{4+}$) form a Kagome lattice. 
Quantum fluctuations will reduce 
the amplitude of charge order substantially. Accompanying charge order, 
we expect some interesting spin liquid phase or complex short range spin 
order at low temperatures}
\end{figure}
Before we consider influence of \wat we discuss some simple charge 
orders at $x = \frac{1}{4}$ and $\frac{1}{3}$ that are favored by 
electrostatics in the unhydrated \naxcob. 
We ignore the superexchange contribution, as $J <<  V_1, V_2$. 
For $x= \frac{1}{4}$, the $Co^{3+}$ sites are arranged on a triangular
lattice (figure 1) to minimize coulomb repulsions. Interestingly, 
the remaining sites carry spins and form spin-\half Heisenberg
antiferromagnet on a Kagome lattice.  In our convention, the classical 
energy of this state is $E_{\frac{1}{4}}(\rm Kagome) = 0$. 
In the real system, carrier delocalization will considerably 
reduce the amplitude of charge order. {\em In this sense the charge ordered
states shown in figure 1 and figure 2 are to be thought of as reference
classical states}.

Another ground state comparable in energy is an anisotropic metal. It has 
ordered stripes - alternating insulating and 0.5 electron doped chains. 
The electrostatic energy of this state per site is $ \frac{1}{4}(V_1 + V_2)$.
However, the carrier delocalization in the 0.5 electron doped chains leads 
to a gain in kinetic energy which is easily estimated when J is 
neglected in our t-J model. This case corresponds to a quarter filled 
infinite U Hubbard model, which can be converted into a half filled band of 
non-interacting spinless fermions giving us the delocalization energy
$ = -|t|\sum \cos k = -2{\frac{|t|}{\pi}}$. Thus we get a total energy
per site, 
$E_{\frac{1}{4}}({\rm stripe}) = \frac{1}{4}(V_1 + V_2) - 
2{\frac{|t|}{\pi}}$.  

Figure 2 shows the case of $x = \frac{1}{3}$. This classical 
ground state minimizes electrostatic repulsion and $Co^{3+}$ sites fill
one of the three sublattices and the remaining hexagonal lattice is
the neutral spin-\half site. It is a hexagonal spin-\half quantum 
antiferromagnet. The energy of this state per site is $E_{\frac{1}{3}} = 
\frac{V_2}{2}$. We also find striped states which are local minima.
\begin{figure}[h]
\epsfxsize 5.5cm
\centerline {\epsfbox{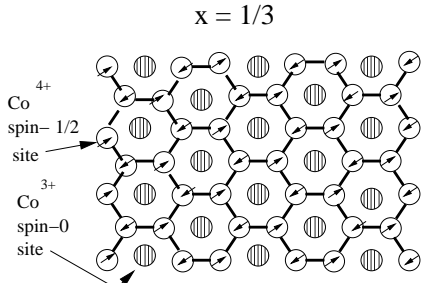}}
\caption{`Classical' charge order for $x = \frac{1}{3}$.
Triangular lattice of localized charge $-e$ carriers and a 
Hexagonal lattice of neutral spin-\half moments. Quantum fluctuations
will reduce the amplitude of charge order substantially. Accompanying
charge order we expect short range AFM order at low temperatures}
\end{figure}

So far we studied charge order in the \cob plane at commensurate
fillings. As we move away from $ x = \frac{1}{3}$ and $ x = \frac{1}{4}$,
defects and discommensurations will be produced or we may go to an 
incommensurate charge ordered structure. There may be one or more 
first order phase boundary between $x = \frac{1}{3}$ and 
$ x = \frac{1}{4}$. However, all these nice charge ordered phases
will be challenged by a generically disordered arrangement of \na 
ions in a triangular lattice and the consequent random potential seen
by the mobile $Co^{3+}$ carriers as explained below . 

The energetically preferred sites of \na atoms\cite{cobstr}
in \naxcob form a triangular lattice that have 
the same lattice parameter as the triangular $Co$ layer. However,
the \na and $Co$ lattices are relatively shifted - if we project
the allowed positions of the \na atoms of the nearest top and 
bottom layer onto the $Co$ layer, these sites become the dual lattice
(hexagonal lattice) of the $Co$ triangular lattice. Because of this
{\em a sublattice order of the \na atoms does not couple to the
charge density wave order-parameter of the \cob lattice and in 
principle allows a finite temperature charge order-disorder 
phase transition}:
a 3-state Potts ($Z_3$ symmetry) model transition at $x = \frac{1}{3}$ 
and a 4-state Potts model ($Z_4$ symmetry) transition at 
$ x = \frac{1}{4}$. However, an inevitable  disorder in \na 
sublattice leads to, based on an Imry-Ma  type of argument, a glassy 
order at low temperatures rather than a genuine charge order phase 
transition. Thus we expect a phase diagram depicted in figure 3 for 
the range $\frac{1}{3} < x < \frac{1}{4}$. 

Let us move on to the hydrated case, \nxcob. As we mentioned 
earlier, the enhanced dielectric constant of the \wat layer
will screen short range coulomb repulsion and weaken and melt 
the high temperature charge ordering. It also screens and weakens
the random \na potential seen by the carriers.
For the appearance of low temperature superconductivity 
$T_{\rm ch}$ need not be reduced to nearly zero value. A 
sufficiently weakened charge ordered state may give up at low 
temperatures and superconductivity may emerge. Figure 3
shows sketches the change of phase diagram as we go to the
hydrated case. A strong resistance anomaly seen in a recent 
experiment\cite{jin}  at $T^* \approx 50~K$, may be 
the weakened charge order transition that we are discussing.
\begin{figure}[h]
\epsfxsize 8.0cm
\centerline {\epsfbox{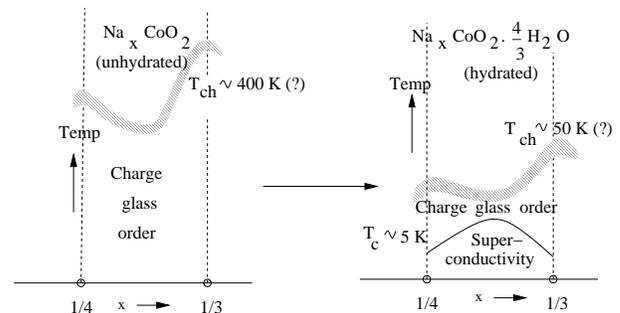}}
\caption{Schematic phase diagram showing how hydration affects 
electronic phases: superconductivity is stabilized at low 
temperatures by a strong suppression of charge glass order.}
\end{figure}
Let us discuss nature of \wat ordering hydrogen bonding 
in \nxcob in some detail. In what follows we use a recent 
suggestion of Cava et al.\cite{cava3} that \wat may have a structure 
similar to the layers in hexagonal ice (1h ice). As mentioned earlier,
in \naxcob, energetically favorable interlayer sites of $Na$ form 
a triangular lattice. These sites are at the center of trigonal 
biprisms, capped by an oxygen atom at the top and one at the bottom. 
To understand \wat ordering, we consider
$Na_{\frac{1}{3}}.CoO_2.{\frac{4}{3}}H_2O$ as a reference compound.
Fill one of of the three sublattices by \na atoms to minimize 
electrostatic energy.  We are left with two empty sublattices that
form a hexagonal lattice. Two \wat molecules may be accommodated at the 
top and bottom of the capped trigonal biprism. By doing so we get two
hexagonal lattices of \wat sandwiching a triangular lattice of \na
ions. Thus we have a triangular lattice filled by \na and \wat in the
ratio $1:4$. We can view the above hexagonal sheets as the sheets in
hexagonal ice structure, as suggested by Cava and collaborators\cite{cava3}
in their
preliminary studies. {\em We find it very interesting that the nearest
neighbor $H_2O$-$H_2O$ distance in the above geometry, $ \approx 2.81~Au$,
is nearly the same\cite{icebook}  as that in real hexagonal ice 
$\approx 2.71$}.
No wonder water may freeze into ice in \nxcob ! Having nearly the same
\wat-\wat distance may also help ice sheets to have a good hydrogen 
bonding network like in hexagonal ice.

Thus it is likely that \wat molecules continue to have hydrogen bonding
in spite of the \cob and \na environment. As hydrogen bond energy is
substantial $\sim 0.5~ eV$, water tends to have hydrogen bonding 
activity in extreme environments. Examples are biological systems, 
clathrate hydrates and water containing charged ions, where \wat 
continue to maintain hydrogen bonding even though the 
local structure deviates considerably from the standard ice or water 
structure. Further the random \na environment in \nxcob may convert
the 2D ice layer into a 2D amorphous ice layer with a good short range 
hexagonal order.

Now we discuss how the dielectric property of the \wat layer may control
the low temperature electronic phases of the conducting \cob layer. We
are interested in finding how the electron-electron interactions 
at the charge order wave vector ${\bf q } = {\bf Q}$ 
get screened by the interacting water dipoles. The relevant static 
dielectric constant is $\epsilon_{\rm H_2O} (Q)$. In an ice system like 
ours with a large disorder in dipole orientations, we expect very small 
variation of $\epsilon_{\rm H_2O}(q)$ with $q$. Further random site 
disorder in the 
\na sublattice will produce Bjerrum defects; so we do not 
expect a 2D dipolar order-disorder transition, as in ideal 2D models 
of ice. 

Dielectric constant of ice has been studied extensively in the past 
and also recently\cite{frohlich,haymet,icebook}. A general expression 
for the dielectric 
constant of an interacting dipolar system is\cite{frohlich,haymet}:
\be
\epsilon = \epsilon_{\infty} + 
\frac{4\pi}{3Vk_BT} \langle ({\bf P} - \langle {\bf P}\rangle )^2 \rangle
\ee
Here $\epsilon_{\infty} \sim 1 - 2$ is the high frequency dielectric 
constant of the dipole, in our case \wat molecule. $\bf P$
is the total dipole moment of the system of volume V. And 
$\langle~...~\rangle$ denotes thermal average.  Static dielectric
constant of ice has not been measured at liquid He temperatures, as the
dielectric relaxation becomes too slow even around liquid air 
temperatures. Fortunately, extensive numerical study of \ewat
are available. For example, a recent calculation\cite{haymet} shows 
that for hexagonal ice, \ewat $\approx 220 $ at $T = 50 K$. 

We use this 3D result to get an approximate estimate for our weakly 
coupled hexagonal ice layers as follows.  We replace the volume 
$V$ by $\approx 6V$ to account for the c-axis expansion
in \nxcob. Missing hydrogen bonds along the c-axis reduces the number
of allowed proton configurations leading to a reduction of 
$\langle ({\bf P} - \langle {\bf P}\rangle )^2 \rangle$ to  
$\approx \frac{2}{3} \langle ({\bf P} - \langle {\bf P}\rangle )^2 \rangle$.
This gives us \ewat (hexagonal sheet) $\approx 20$ at $T = 50 K$. Our
system being strongly disordered, we do not expect \ewat (hexagonal sheet)
to change at lower temperatures. Thus the background dielectric constant
of \nxcob is $\epsilon = \epsilon_o + \epsilon_{\rm H_2O} \approx
8 + 20$. This reduces the mean short range repulsion by nearly a factor
of 3, making the $T_{\rm ch} \approx 1 K$, in equation (2). Once the 
long range charge order
is disabled by a reduction of V, the simple t-J model and consequent
low temperature superconducting phase is realized, albeit with a reduced
\tc in the range $\frac{1}{3} < x < \frac{1}{4}$. The sharp reduction
in superconducting \tc at the commensurate boundaries of the dome should 
arise from the strong short range order and lesser discommensurations and
defects. The experimentally seen flat value of \tc $\approx 2 K$ 
for $x < \frac{1}{4}$ and for $x > \frac{1}{3}$is likely to be an effect 
of phase separation.
\begin{figure}[h]
\epsfxsize 6.0cm
\centerline {\epsfbox{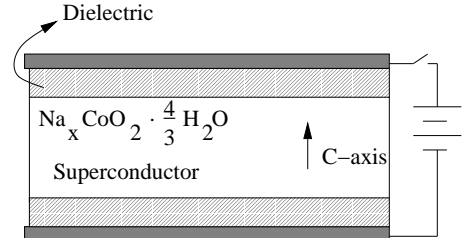}}
\caption{Schematic experiment to observe the `Electrical  
Modulation of Superconducitivy'. The order voltage required
is $500~V$, when the film thickness is about a micron.}
\end{figure}
Our proposal of a critical and catalytic role of \wat layer suggests
ways to access and control low temperature electronic phases of the 
\cob layer. Based on this we suggest two effects: i) `Electrical 
Modulation of Superconductivity'. Here we control (figure 4) 
superconductivity 
and superconducting \tc by modifying the screening property of \wat
layer by external electric fields - DC, AC or pulsed fields. 
This has interesting consequences of 
being able to locally erase superconductivity by STM tips, dynamically
create Josephson networks or create 2D superconductivity of desired
shapes through appropriate capacitor shapes etc. 
Since microwaves are absorved by hydrogen bonded networks we can pump 
microwaves at appropriate frequencies ($\hbar \omega <
\Delta_{\rm sc}$, the superconducting gap) and dynamically polarize 
water dipoles and may influence its dielectric properties, and in turn 
control superconductivity.

ii) `Electroresistance Effect'. By modifying the amplitude of 
charge glass order as well as $T_{\rm ch}$ in the 
non-superconducting state, by influencing the \wat layer by
external DC or AC electric field,
we can change $\rho_{ab}$, the ab-plane resistivity. 

Below we estimate the electric field required to completely suppress 
superconductivity. Having established a hydrogen bonded
network it requires a finite energy to rotate a water molecule and
orient its dipole moment along an external electric field. In infrared 
absorption
and neutron scattering experiments\cite{icebook}
the absorption band corresponding to 
rotation of \wat molecules is in the range $60~{\rm to}~120~meV$. 
Assuming a random orientation, the average energy required to reorient 
a water molecule is $\approx 50~meV$, i.e., a potential of
$50~{\rm mV}$ applied over the length $\approx 1~Au$ of the
water dipole will orient the dipole moment of water along its field.
If we have c-axis oriented \nxcob film of thickness $1$ micron we need
to apply a voltage $\approx 500~Volts$ across the film, in order to 
orient the majority of dipoles. Strong polarization of water dipoles 
reduces the dielectric constant, as is evident from equation (3).
The resulting reduced screening of carriers in the \cob layer
allows charge order to grow and superconductivity gets suppressed.

A theoretical analysis, including some of the possible difficulties 
in observing the effects will be presented in a future publication. 

To get a clear understanding of this complex system, and to see
if our proposal is correct more experiments are necessary:

a) {\bf Charge Order:} It will be interesting to perform NMR,
NQR, STM, $\mu SR$ and other local probe measurements to look for 
charge order in the vicinity of the commensurate fillings 
$x = \frac{1}{4}, \frac{1}{3}, \frac{1}{2}, \frac{2}{3}$ and 
$\frac{3}{4}$ and see how they differ between the two systems
\nxcob and \naxcob. 

b) {\bf Spin order, singlets and gaps:} Accompanying local charge order 
we expect a spin order at 
low temperatures (the scale of J is small, $\sim 6$ to $7~meV$). 
In general the enhanced singlet stabilization by the superexchange
process will introduce some kind of spin gap phenomenon. If the charge 
order at $x = \frac{1}{4}$ leads to a Kagome lattice of spins it will be 
an interesting testing ground for some of the ideas of the spin
liquid phase of spin-\half Kagome antiferromagnet, including possible
novel excitations.

c) {\bf Lower Doping:} Experimentally, it has not been 
possible\cite{cava2}
to make \nxcob for $x < \frac{1}{4}$. It is likely\cite{cava2} that 
c-axis ionic bonding is weakened, by decreasing $x$ and 
presence of water layer, making a 3D structure unstable.
It will be important to synthesize, by 
non-equilibrium means, meta stable compounds for $x < \frac{1}{4}$
to test the validity of RVB theory and also test our hypothesis
of the role played by water.

d) {\bf Replacing \wat:} It will be desirable to have a stable solid
$Na_x CoO_2.y X$, where an intercalant `X' not only increases the 
dielectric constant but also provides additional bonding between 
\cob layers and make stable compounds for $x < \frac{1}{4}$.

e) {\bf Higher doping:} According to reference 6, the dopant 
induced dynamics, within the t-J model will favor ferromagnetic 
correlations and a consequent p-wave superconductivity at higher 
dopings slightly above $x = \frac{1}{3}$. It will be interesting 
to look for this.

f) {\bf Inhomogeneous Superconductivity:} If ice plays a central
role, as suggested in this paper, \wat density fluctuation in  
ice layer will directly influence superconductivity in nearby 
\cob layers resulting in a corresponding fluctuation in the 
superconducting order parameter and possible well grown local charge 
ordered phase. 

g) {\bf Slow Relaxation:} Since interacting water dipoles have very 
slow dielectric relaxation time scales\cite{icebook}, they may 
consequently affect superconductivity and impose some anomalous 
relaxation/aging effects. 

The present paper is phenomenological and qualitative in character.
Any detailed quantitative calculations of \tc and phase diagram
for this complex system needs further experimental guidance.
Issue of calculating local screening and dielectric constant in 
hydrogen bonded systems is known to have subtleties; added to 
this, we have conducting layers sandwiching water layers. We have 
made very crude estimates based on simple physical arguments and
very approximate considerations, as our primary aim is to focus 
and identify how water could play a crucial role in this complex 
system.

We thank A.K. Mishra, V.N. Muthukumar, Debanand Sa, Manas Sardar 
and R. Shankar for discussion and Latha Malar Baskaran for 
reference [18].

\end{multicols}
\end{document}